\documentclass[conference]{IEEEtran}
\IEEEoverridecommandlockouts
\usepackage{color}
\usepackage{cite}
\usepackage{amsmath,amssymb,amsfonts}
\usepackage{algorithmic}
\usepackage{graphicx}
\usepackage{textcomp}
\usepackage[utf8]{inputenc}
\usepackage[T1]{fontenc}
\usepackage{times} 
\usepackage{flushend}
\usepackage{fancyhdr} 
\usepackage{kantlipsum} 
\usepackage{eso-pic} 
\pdfoutput=1
\begin{document}

\title{Development of a Social Network for Research Support and Individual Well-being Improvement}
\author{\IEEEauthorblockN{Lucas V. A. Caldas}
\IEEEauthorblockA{\textit{Center of Behavioral Theory and Research} \\
\textit{Federal University of Pará}\\
Belém, Brazil \\
lvinicius123@gmail.com}
\and
\IEEEauthorblockN{Antonio F. L. Jacob Jr.}
\IEEEauthorblockA{\textit{Technology Sciences Center} \\
\textit{State University of Maranhão}\\
São Luís, Brazil \\
jacobjr@engcomp.uema.br}
\and
\IEEEauthorblockN{Simone S. C. Silva}
\IEEEauthorblockA{\textit{Center of Behavioral Theory and Research} \\
\textit{Federal University of Pará}\\
Belém, Brazil \\
symon@ufpa.br}
\and
\IEEEauthorblockN{Fernando A. R. Pontes}
\IEEEauthorblockA{\textit{Center of Behavioral Theory and Research} \\
\textit{Federal University of Pará}\\
Belém, Brazil \\
farp1304@gmail.com}
\and
\IEEEauthorblockN{Fábio M. F. Lobato}
\IEEEauthorblockA{\textit{Engineering and Geoscience Institute} \\
\textit{Federal University of Western Pará}\\
Santarém, Brazil \\
fabio.lobato@ufopa.edu.br}
}
\maketitle

\fancyhf{} 
\fancypagestyle{plain}{ 
\fancyhf{} 
\fancyhead[C]{Conference on \LaTeX} 
\renewcommand{\footrulewidth}{0pt} 
} 

\IEEEpubid{\parbox{\columnwidth}{\vspace{30pt}{Accepted at the 2018 IEEE/ACM International Conference on Advances in Social Networks Analysis and Mining (ASONAM 18). For academic
purposes, please cite the conference version.}  \hfill} \hspace{\columnsep}\makebox[\columnwidth]{}} 
\IEEEpubidadjcol 

\begin{abstract}
The ways of communication and social interactions are changing drastically. Web users are becoming increasingly engaged with Online Social Networks (OSN), which has a significant impact on the relationship mechanisms between individuals and communities. Most OSN platforms have strict policies regarding data access, harming its usage in psychological and social phenomena studies; It is also impacting the development of computational methods to evaluate and improve social and individual well-being via the web. Aiming to fill this gap, we propose in this paper a platform that brings together social networks dynamics with forum features, altogether with gamification elements, targeting researchers interested in obtaining access to user's data their investigations.
\end{abstract}


\section{Introduction}
\label{sec:introduction}

Web users are becoming increasingly engaged with Online Social Networks (OSN). Thus, it is perceived a significant expansion of OSN applications boundaries to fields such as academic production, product reputation, human resources management, among others \cite{Lobato2017}. This phenomenon is related to the spread of Web 2.0, which has consolidated the presence of systems enabling people´s interaction, information sharing, and group formation.  Web users are not anymore just information consumers, once they produce and consume content at the same time, named as ``\textit{prosumers}'' \cite{prosumer}. As a consequence, the social capital concept is gaining even more notoriety, which can be defined as a form of economic and cultural capital where social networks play a central role, and in which transactions are marked by reciprocity, trust, and cooperation among agents \cite{Adler2012}.

In this context, social support via virtual environments arises as users actively post content, get connected with other people and exchange experiences \cite{Smedley:2016, SMEDLEY201553}. This support is directly related to the amount of help in those environments. Besides, it can even be obtained through social relationships to meet different types of need and to support those who are receiving it \cite{Rand2011}. In well-established OSNs such as Facebook, social support is provided through communities and direct interactions. However, these communities are often private and have access restrictions. Moreover, it should be noted that, despite the pervasiveness of such platforms, well-established OSN have severe restrictions on access to data due to privacy and information security policies \cite{Lobato2017}. Such policies make it difficult to use the user-generated content in social science studies, thus, impacting negatively on some understanding of social phenomena.

In the light of the foregoing considerations, an \textit{International Research Consortium in Social Networks Analysis} stated that they need a platform which: i) provides desired features to support social science studies, ii) minimizes access restrictions to user-generated content for academic researchers, and, at the same time, iii) provides an environment to improve individual well-being by inducing the communities formation, and, consequently, increasing social capital. Aiming to fill this gap, we present in this paper a platform that mixes network dynamics with forums and gamification elements. It is designed for researchers interested in obtaining access to user data for academic studies, as well as for people interested in sharing their experiences and in building up relationships for mutual support and cooperation. The platform was developed from the perspective of \textit{Design Science Research} (DSR). Given these aspects, the following Research Questions (RQ) were raised:

\begin{itemize}
    \item RQ1:  Is it possible to design a social network platform to support  studies on social and psychological phenomena?
    \item RQ2: What is the impact of such platform in the users´ well-being?
\end{itemize}

The remainder of this paper is organized as follows. Section \ref{sec:social}  presents the background of this paper, discussing social networks analysis and its importance to social science studies. The proposed platform is described in Section \ref{sec:Result}. Finally, Section \ref{sec:FinalRemark} concludes this paper and presents future research directions.

\section{Social Networks in Social Sciences Researches}
\label{sec:social}

Personal and social networks are essential for exchanging experiences and inducing mutual assistance among their participants. \cite{Donmus2010} defines a social network as a system composed of several participants, which have functions and activities applied to an interaction context. In other words, it is a set of actors (individuals) and their connections (relationships). These networks can be categorized considering the environment where social interactions occur: in-person or virtual environment, for instance. In this sense, groups and communities are formed in a way that there are information sharing and people integration \cite{Wei2012, Silva2017}. These are peculiar features existing in social support networks, where interactions tend to be mutually beneficial to its participants, and may also be the result of the friendship acquisition \cite{Thi2017}.

The lack of social capital is common to people who experience situations that lead them away from social environments, which is the case of relatives of people with disabilities \cite{Hendricks2015}. In several areas, it is observed that the induction of a social support network in OSN enables a more pervasive and effective formation of virtual communities for mutual support \cite{OH201469}.  The supportive interactions in users groups can be predicted through logs analysis of systems usage. For instance, \cite{Glenski} has analyzed posts, clicks and page ladings made by Reddit users. As a result, the authors have measured what they called ``interaction success'' of users posts. This information can be used to recommend posts/content that is more relevant to each kind of user.

In another point of view, \cite{Studies2017} has analyzed an in-person social network, modeling it as a virtual environment. To do so, the authors modelled a network structure to identify the public participation in the management of protected areas, such as the Iron Gates Natural Park (Romanian). The data collected allowed the authors to build a graph corresponding to the social network interactions. As a result, the authors perceived that the networks for protected area management have weak connections between its nodes, and that they are very decentralized. Thus, this finding has justified the necessity to improve the collaboration between the people involved in this process.  In other perspective, \cite{Hendricks2015} studied adolescent mothers from Western Australia. The authors became aware that the easy access to information through existing communities in virtual social networks has contributed to the increase in the sense of trust, competence, and responsibility. Also, the use of networks softens the sense of social isolation through the re-engagement of these users in their social circles.  

These are just a few examples of virtual network analysis and its relevance for social science studies. Even though, many more applications could be targeted in this context if researchers could have full access to the user-generated content and related data. In this sense, an \textit{International Research Consortium in Social Networks Analysis}, formed by research centers in Sociology, Psychology, Nursing, Economics and Computing, stated a necessity for a platform with free and complete access to data, considered crucial to increase the impact of the actions and studies conducted. This consortium is composed by several research groups and laboratories, with international cooperation, the main partners in Brazil are:  The Federal University of Pará is represented by the Computational Intelligence Laboratory\footnote{http://www.linc.ufpa.br/} and the Development Ecology Laboratory\footnote{http://www.ufpa.br/led/}, the latter involves researchers from psychology, occupational therapy and sociology; The NeuroRehab group is representing University of São Paulo\footnote{http://gruposdepesquisa.eerp.usp.br/sites/neurorehab/}; The Federal University of Minas Gerais is participating with the  Interdisciplinary Research Group in Social Network Analysis\footnote{http://www.giars.ufmg.br/}; and the Federal University of Western Pará, through the Applied Computing Research Group\footnote{dgp.cnpq.br/dgp/espelhogrupo/2969843796195905}. The major international cooperation came from the Social CRM Research Center of the Universit{\"a}t Leipzig\footnote{http://www.scrc-leipzig.de/} and the Center for Counseling and Therapy from the Universit{\"a}t Dortmund\footnote{https://www.fk-reha.tu-dortmund.de/zbt/de/home/index.html}. Some of the investigations conducted by this consortium include the use of social networks (virtual and face-to-face) for parental support to families with relatives with disabilities; innovation and collaboration ecosystems; and social customer relationship management OSN. In the light of the given scenario, the next section presents the proposed platform.

\section{Results}
\label{sec:Result}

\begin{figure}[!ht]
    \centering
    \includegraphics[width=0.47\textwidth]{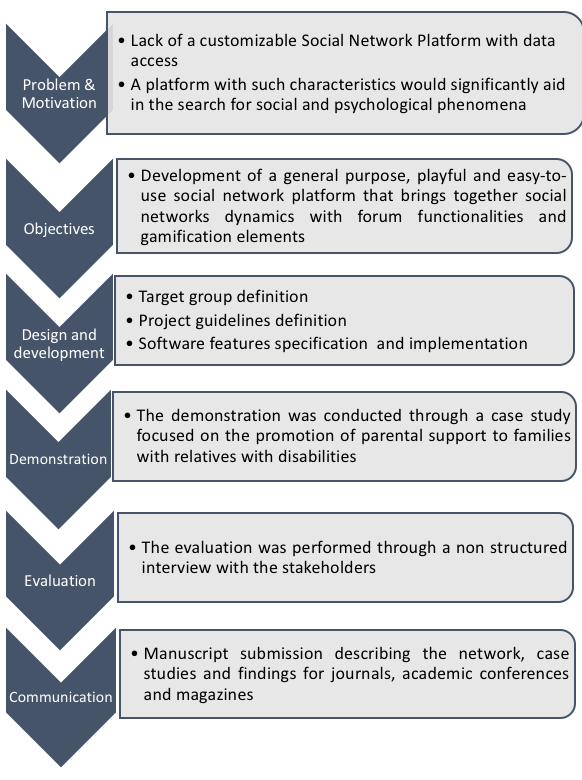}
    \caption{Instantiation of the DSR elements of the development of the proposed platform.}
    \label{fig:dsr}
    \end{figure}

The artifact was developed under \textit{Design Science Research} methodology, which is a well-accepted approach in information systems research projects when the final goal is the development of a product, strategies, and services. According to \cite{Peffers:2007}, the DSR is a process model composed of six sequential steps, which are: i) problem identification and motivation; ii) definition of solution objectives; iii) project and development; iv) demonstration; v) evaluation; and vi) communication. The Figure \ref{fig:dsr} shows the DSR elements for the present project. 

The aforementioned research consortium required a platform that would provide features usually not found in well-established social networks. Considering that the system should have the flexibility to be applied in different contexts, the project guidelines were defined in meetings with stakeholders from the consortium to ensure the attendance of all their needs. So, the following guidelines were established: 1) ease of navigation; 2) responsiveness; 3) information security; 4) flexibility and extensibility; and 5) user engagement.

The architecture of the network developed was based on the Model-View-Controller (MVC) pattern, to meet the guideline 3. Guidelines 1 and 2 were used and applied in the View layer, while the others were implemented in the Model layer because they were business rules or non-functional security requirements (guideline 3). There was also a concern with bandwidth consumption, considering that a significant part of the users would access the platform by mobile devices. It has also motivated the implementation of some strategies to decrease bandwidth consumption in the controller layer. The platform was developed in PHP programming language, using the CodeIgniter framework, the database was implemented using MySQL and the responsiveness was implemented using Bootstrap. Finally, to ease the user experience, the system was inspired by popular networks such as Facebook, Twitter, and Orkut. 

Thereby, many of the proposed platform features have been inspired by features that already exist in other networks, but with several elements mixed, providing a differential from the already existing systems. For instance, one of these unique features is the possibility of posting PDF documents. 
In Table \ref{tab:comparativo} we can see a comparative overview of the main features of the proposed system and some of the most used networks.

\begin{table}[!ht]
\centering
\caption{Comparisons between main functionalities of the most used networks and the proposed network.}
\label{tab:comparativo}
\begin{tabular}{|c|c|c|c|c|c|}
\hline
\textbf{Functionalities}             & \rotatebox[origin=c]{80}{\quad Proposed System\quad} & \rotatebox[origin=c]{80}{Facebook} & \rotatebox[origin=c]{80}{Twitter} &\rotatebox[origin=c]{80}{Youtube} & \rotatebox[origin=c]{80}{Whatsapp} \\
    \hline

\textbf{\begin{tabular}[c]{@{}c@{}}Open  Login\end{tabular}}              & \textcolor{green}{\checkmark }                                                         & \textcolor{green}{\checkmark }       & \textcolor{green}{\checkmark }      & \textcolor{green}{\checkmark }      & \textcolor{green}{\checkmark }       \\ \hline
\textbf{\begin{tabular}[c]{@{}c@{}}Create  Communities\end{tabular}}       & \textcolor{green}{\textcolor{red}{$\times$}}                                                            & \textcolor{green}{\checkmark }       & \textcolor{green}{\textcolor{red}{$\times$}}         & \textcolor{green}{\textcolor{red}{$\times$}}         & \textcolor{green}{\checkmark }       \\ \hline
\textbf{Survey}                                                             & \textcolor{green}{\checkmark }                                                         & \textcolor{green}{\checkmark }       & \textcolor{green}{\textcolor{red}{$\times$}}         & \textcolor{green}{\textcolor{red}{$\times$}}         & \textcolor{green}{\textcolor{red}{$\times$}}          \\ \hline
\textbf{Gamification}                                                       & \textcolor{green}{\checkmark }                                                         & \textcolor{green}{\textcolor{red}{$\times$}}          & \textcolor{green}{\textcolor{red}{$\times$}}         & \textcolor{green}{\textcolor{red}{$\times$}}         & \textcolor{green}{\textcolor{red}{$\times$}}          \\ \hline
\textbf{\begin{tabular}[c]{@{}c@{}}Profile Customization\end{tabular}}   & \textcolor{green}{\checkmark }                                                         & \textcolor{green}{\checkmark }       & \textcolor{green}{\checkmark }      & \textcolor{green}{\checkmark }      & \textcolor{green}{\checkmark }       \\ \hline
\textbf{Chat}                                                               & \textcolor{green}{\checkmark }                                                         & \textcolor{green}{\checkmark }       & \textcolor{green}{\textcolor{red}{$\times$}}         & \textcolor{green}{\textcolor{red}{$\times$}}         & \textcolor{green}{\checkmark }       \\ \hline
\textbf{Video}                                                              & \textcolor{green}{\checkmark }                                                         & \textcolor{green}{\checkmark }       & \textcolor{green}{\checkmark }      & \textcolor{green}{\checkmark }      & \textcolor{green}{\checkmark }       \\ \hline
\textbf{\begin{tabular}[c]{@{}c@{}}Post  Documents (PDF)\end{tabular}}      & \textcolor{green}{\checkmark }                                                         & \textcolor{green}{\textcolor{red}{$\times$}}          & \textcolor{green}{\textcolor{red}{$\times$}}         & \textcolor{green}{\textcolor{red}{$\times$}}         & \textcolor{green}{\checkmark }       \\ \hline
\textbf{Events}                                                             & \textcolor{green}{\textcolor{red}{$\times$}}                                                            & \textcolor{green}{\checkmark }       & \textcolor{green}{\textcolor{red}{$\times$}}         & \textcolor{green}{\checkmark }      & \textcolor{green}{\textcolor{red}{$\times$}}          \\ \hline
\textbf{Lives}                                                              & \textcolor{green}{\textcolor{red}{$\times$}}                                                            & \textcolor{green}{\checkmark }       & \textcolor{green}{\textcolor{red}{$\times$}}         & \textcolor{green}{\checkmark }      & \textcolor{green}{\textcolor{red}{$\times$}}          \\ \hline
\textbf{\begin{tabular}[c]{@{}c@{}}Profile Restrictions\end{tabular}}    & \textcolor{green}{\textcolor{red}{$\times$}}                                                            & \textcolor{green}{\checkmark }       & \textcolor{green}{\checkmark }      & \textcolor{green}{\checkmark }      & \textcolor{green}{\checkmark }       \\ \hline
\textbf{\begin{tabular}[c]{@{}c@{}}Unrestricted  Data Acess\end{tabular}} & \textcolor{green}{\checkmark }                                                         & \textcolor{green}{\textcolor{red}{$\times$}}          & \textcolor{green}{\textcolor{red}{$\times$}}         & \textcolor{green}{\textcolor{red}{$\times$}}         & \textcolor{green}{\textcolor{red}{$\times$}}          \\ \hline
\end{tabular}
\end{table}

There are two types of users´ on the platform:  Moderators and Ordinary users. Moderators are users with special privileges. For example, they are responsible for managing the network communities.  Another point that deserves attention is the merging of social networking elements with forums attributes. This concept was integrated into the platform through a discussion system inside the communities. 

Therefore, communities were designed to work as chat groups, where each community represents a subject that can be divided into smaller and more specific discussions. For instance, in a system development context, we could have the following communities: ``Programming Language'', ``Database''  and ``Software Engineering´´. Thus, discussions are smaller topics within a community subject and are also the place where users interact with each other to talk about that topic. In the ``Programming Language'' community, we could have discussions named ``PHP'', ``JAVA'' and ``C++''. 

Ordinary users are not allowed to create new communities, once each community needs a moderator, and this function is only given to a person who detains some knowledge about the community subject. However, any user may create discussions, and by doing so, they propose new themes and keep the community alive. Communities also have a survey application system, which works as a quick way of understanding the opinion of members through small questions with multiple choice and predefined answers. This mechanism was added to support researchers in survey applications.

Thinking about encouraging users participation in these questions, they were incorporated into the platform gamification mechanism, also to meet guideline 5. Software gamification is a technique that seeks to reduce failures, increase functionality and engagement. However, misapplication and lack of intimacy with the gamification process design can make it a big enemy, leading to very high risks for a software that may lead to its failure \cite{Technology2017, Deterding2017}. The main gamification elements adopted were points, levels, rankings, medals/achievements, integration, engagement loops, and reinforcement and feedback rules. These elements have unique characteristics and distinct purposes. Among these, we highlight the engagement loops, which are bonuses and missions that arouse the user's interest in reusing the system, which is directly related to a motivational feeling.

The gamification elements were applied to the network structure through a reward system based on actions performed by users, such as liking and sharing posts, answering surveys and posting activities. Each action is considered to raise the user level, and each level unlocks a medal. The difficulty of obtaining the next medals is computed as an arithmetic progression, that always adds twice the amount of experience to go to the next level. The network has a total of nine medals. 

Some desired features such as "lives" and "events" were not added yet. However, strategies for inclusion of these features are already being studied. Besides that, it is worth mentioning that the unrestricted data access is available only to the researchers who use the platform as a tool for research support and that sign a compromise term. The platform terms of use describe that user data can be used in scientific studies. By accepting it, the user is granting permission to use their data in such research projects.       

\section{Final Remarks}
\label{sec:FinalRemark}
Online Social Networks had revolutionized social science studies. These platforms represent a new communication channel and a relevant data source about individual's behavior and desires. However, those platforms lack some features such as community creation, possibility to apply surveys or to share documents, for instance. Moreover, most large-scale studies face a critical barrier regarding restrictions on data access. In the light of the foregoing considerations, an International Research Consortium stated a need regarding a platform that provides desired features to support social science studies and minimizes access restrictions to user-generated content for academic researchers. 

In this paper, we have tackled these issues by developing a general purpose social network that brings together social networks dynamics, forum features and gamification elements; targeting researchers interested in obtaining access to users´ data to study psychological and social phenomena. To do so, Design Science Research was adopted as a methodology to conduct the present study. 

Retaking the research questions stated before, and analyzing the results, it is possible to answer them. For the RQ1, considering the preliminary results and the researchers' expectations, the answer is yes, it is possible to use such kind of platform. Besides, its usage is not restricted only to academia, but also able to be applied to industry as well. One application glimpsed was to promote innovation ecosystems in medium companies. Regarding RQ2, the platform features, more specifically the forums with moderator and the gamification strategies, have helped to improve the information quality across the network, but more analysis should be performed.

Considering these results, for future work, more case studies should be conducted and quantitative evaluation should be performed to assess researcher´s and volunteer´s satisfaction.

\section*{Acknowledgements}
\label{sec:Acknowledgements}
This research was supported by the National Council for the Improvement of Higher Education (CAPES); the National Council for Scientific and Technological Development (CNPq); and Amazon Foundation for Studies and Research (Fapespa); all of them are Brazilian.

\bibliographystyle{IEEEtran}
\bibliography{conference_071817}

\end{document}